\documentclass[reqno]{amsart}
\usepackage{amsmath,amssymb,euscript,upgreek,graphicx}
\usepackage[arrow,matrix,curve]{xy}
\numberwithin{equation}{section}

\newcommand\bZ{\mathbb{Z}}
\newcommand\bN{\mathbb{N}}
\newcommand\bR{\mathbb{R}}
\newcommand\bC{\mathbb{C}}
\newcommand\cB{\EuScript{B}}
\newcommand\cH{\EuScript{H}}
\newcommand\cJ{\EuScript{J}}
\newcommand\cK{\EuScript{K}}
\newcommand\cP{\EuScript{P}}
\newcommand\cV{\EuScript{V}}
\newcommand\Gr{\mathrm{Gr}}
\newcommand\SO{\mathrm{SO}}

\newcommand\Sp{\mathrm{Sp}}
\newcommand\U{\mathrm{U}}
\newcommand\Hom{\operatorname{Hom}}
\newcommand\tr{\operatorname{tr}}
\newcommand\id{\operatorname{id}}
\newcommand\dr{\operatorname{d}\!}
\newcommand\pdr{\partial}
\newcommand\ii{\sqrt{-1}}
\newcommand\bra\langle
\newcommand\ket\rangle
\newcommand\medwedge{\scalebox{1}[1.25]{$\wedge$}\hspace{1pt}}
\newcommand\gam\gamma
\newcommand\del\delta
\newcommand\eps\epsilon
\newcommand\sig\sigma
\newcommand\tht\theta
\newcommand\om\omega
\newcommand\upsig\upsigma
\newcommand\Gam\varGamma
\newcommand\Om\varOmega
\newcommand\tdr{\tilde{\operatorname{d}}}
\newcommand\tS{{\tilde S}}
\newcommand\Tv{T^{\mathrm{v}}}
\newcommand\ab{_{ab}}
\newcommand\mfrac[2]{\raisebox{-.1ex}{\scalebox{1.1}[1.15]{$\frac{#1}{#2}$}}}

\begin{document}

$\;$

\bigskip

\begin{center}
{\huge\bf Quantisation of a Family of Phase Spaces}

\vspace{3ex}

{\large\bf Siye Wu}\footnote{Department of Mathematics,
National Tsing Hua University, Hsinchu 30013, Taiwan.\\
\hspace*{4.5ex}E-mail address: {\tt swu@math.nthu.edu.tw}}
\end{center}

\bigskip
\begin{flushright}
{\small\em Dedicated to Armen Sergeev on the occasion of his 70th birthday}
\end{flushright}
\bigskip

\begin{quote}
{\bf Abstract}---We explain that when quantising phase spaces with varying
symplectic structures, the bundle of quantum Hilbert spaces over the parameter
space has a natural unitary connection.
We then focus on symplectic vector spaces and their fermionic counterparts.
After reviewing how the quantum Hilbert space depends on physical parameters
such as the Hamiltonian and unphysical parameters such as choices of
polarisations, we study the connection, curvature and phases of the Hilbert
space bundle when the phase space structure itself varies.
We apply the results to the two-sphere family of symplectic structures on a
hyper-K\"ahler vector space and to their fermionic analogue, and conclude with
possible generalisations.
\end{quote}

\bigskip

\section*{Introduction}
The classical phase space of a bosonic particle is a symplectic manifold.
Quantisation aims to obtain a Hilbert space on which classical observables
act as operators.
In geometric quantisation, the Hilbert space depends on the symplectic
structure, the prequantum line bundle as well as the choice of a polarisation,
which is usually regarded as unphysical data.
There is a bundle of Hilbert spaces over the space of polarisations.
Projective flatness of this bundle means that the notion of quantum states
is independent of the choice of the polarisation.
But this is not usually achieved unless the phase space is a symplectic vector
space or its reduction, and the polarisations are restricted to linear ones.
In this paper, we assume that polarisations are already suitably chosen but
concentrate mostly on the quantisation of phase spaces with varying physical
parameters, such as symplectic forms and Hamiltonians, and we study the
geometry of the bundle of quantum Hilbert spaces.
Holonomies of this bundle are geometric phases that have physical relevance.

The article is organised as follows.
In \S\ref{gen}, we recall the basic theory of geometric quantisation and
explain how the quantum Hilbert space depends on the choices of a prequantum
line bundle and a polarisation.
When there is a family of symplectic manifolds, we construct a connection on
the bundle of Hilbert spaces and explain the resulting (possibly non-Abelian)
geometric phases.
In \S\ref{lin}, we specialise to linear phase spaces.
If the symplectic form is fixed, we review the projectively flat connection
(flat if metaplectic correction is used) on the bundle of Hilbert spaces
over the space of linear polarisations.
If the symplectic structure is varying in a suitable family, we fix a suitable
polarisation for each symplectic structure and describe explicitly the
connection, curvature and phases on the bundle of Hilbert spaces.
We also consider the same problems for linear fermionic phase spaces, which
have Euclidean rather than symplectic structures.
We then consider quantisation of a hyper-K\"ahler vector space with a family
of symplectic structures parametrised by the $2$-sphere.
The fermionic analogue is given by a paraquaternionic K\"ahler structure.
We conclude by summarising the significance of our examples of linear phase
spaces and speculate on the generalisations to curved hyper-K\"ahler or
paraquaternionic K\"ahler manifolds.

\section{General framework}\label{gen}

\subsection{Quantisation of a single phase space}\label{gen-fix}
We briefly recall the standard theory of geometric quantisation to set up
and fix the notations.
For details, see the original treatises \cite{Ko,So} or textbooks and survey
articles \cite{Bl,Sn,Ki}.

The phase space of a classical bosonic system is a symplectic manifold
$(M,\om)$.
A prequantum line bundle over $M$ is a line bundle $\ell$ with a unitary
connection $\nabla^\ell$ whose curvature is $\om/\ii$.
As such its first Chern class $c_1(\ell)\in H^2(M,\bZ)$ modulo torsion is
represented in the de Rham cohomology by the closed $2$-form $\om/2\pi$.
With $\om$ fixed, the torsion part of $c_1(\ell)$ can be affected by tensoring
$\ell$ by flat line bundles over $M$, whose first Chern classes form the
torsion subgroup of $H^2(M,\bZ)$.
The space of topologically trivial flat line bundles can be identified with
$H^1(M,\bZ)\otimes\U(1)$ (cf.~\cite{FMS}) and is a connected component of the
space $H^1(M,\U(1))=\Hom(\pi_1(M),\U(1))$ of all flat line bundles up to gauge
equivalences.
With a fixed $c_1(\ell)$ or fixed topology, the remaining ambiguity in $\ell$ 
can be absorbed by adding to the connection a harmonic $1$-form (modulo gauge
transformations) that represents an element in $H^1(M,\bZ)\otimes\U(1)$.
Henceforth we assume that a prequamtum line bundle $\ell$ over $(M,\om)$ with
a connection $\nabla^\ell$ is chosen.

Quantisation assigns a quantum Hilbert space to a symplectic phase space
$(M,\om)$ with a prequantum line bundle $\ell$.
Technically, this requires choosing a (non-negative) polarisation $P$, a
complex Lagrangian distribution on $M$ such that $-\ii\,\om(X,\bar X)\ge0$
for all $X\in\Gam(P)$.
For example, $P_J:=T^{1,0}_JM$ is a polarisation if $J$ is an almost complex
structure on $M$ compatible with $\om$, i.e.,
$\om(J\cdot,J\cdot)=\om(\cdot,\cdot)$ and $\om(\cdot,J\cdot)$ is positive
definite.
The Hilbert space $\cH_J$ is then the $L^2$-completion of the space of
sections of $\ell$ that are covariantly constant (or parallel) along
$\bar P_J=T^{0,1}_JM$;
the norm is from the Hermitian inner product defined by the pointwise pairing
of two sections followed by an integration over $M$ using the Liouville volume
form $\om^n/n!$.
With the metaplectic or half-form correction, which shifts the integrality
condition on $\om/2\pi$ by $\frac12c_1(K_P)$, we consider instead sections of
$\ell\otimes\sqrt{K_P}$, where
$K_P:=\medwedge^\mathrm{\!top}(T^\bC M/\bar P)^*$ is the canonical line bundle
of $P$.
There is a partial connection along $(0,1)$-vectors on half-forms or on
$\sqrt{K_P}$; for example, if the polarisation $P=P_J$ comes from an almost
complex structure $J$, then $T^\bC M/\bar P\cong P$ and $K_P$ is the canonical
line bundle over $(M,J)$, and we can use the (unique) connection on $TM$
preserving both $\om$ and $J$ with torsion $\frac14N_J$, where $N_J$ is the
Nijenhuis tensor of $J$.
The metaplectically corrected quantum Hilbert space $\hat\cH_P$ is the space
of sections of $\ell\otimes\sqrt{K_P}$ that are covariantly constant along
$\bar P$, completed by an $L^2$-norm explained below.

For more generality, we assume that there are two polarisations $P$ and $P'$
such that the rank of $\bar P\cap P'$ is constant.
Then there are two real distributions $D$ and $E$ on $M$ such that
$\bar P\cap P'=D^\bC$ and $\bar P+P'=E^\bC$.
We further assume that $E$, and hence also $D=E^\om$, is involutive and that
the $D$-leaves in $M$ form an orientable manifold $M/D$.
The symplectic form $\om$ defines an isomorphism
$\sqrt{K_{\bar P}}\otimes\sqrt{K_{P'}}\cong\medwedge^\mathrm{\!top}(TM/D)^\bC$.
So if $\psi$ and $\psi'$ are sections of $\ell\otimes\sqrt{K_P}$ and
$\ell\otimes\sqrt{K_{P'}}$ that are parallel along $\bar P$ and $\bar P'$,
respectively, the pointwise pairing $\bar\psi\otimes\psi'$ on $M$ as a section
of $\medwedge^\mathrm{\!top}(TM/D)^\bC$ descends to a top form on $M/D$ which
can be integrated to yield the BKS (Blattner-Kostant-Sternberg) pairing
$\bra\psi,\psi'\ket_{\mathrm{BKS}}$.
If $P=P_J$ for an almost complex structure $J$, then we have $D=\{0\}$ and the
integration is over $M$.
In particular, if $P'=P$ and it satisfies the above conditions, then there is
a norm $\|\cdot\|_P^2=\bra\cdot,\cdot\ket_P$ on the space
$\Gam_{\bar P}(\ell\otimes\sqrt{K_P})$ of sections being parallel along
$\bar P$, and the $L^2$-completion leads to the Hilbert space $\hat\cH_P$.
More generally, we have the BKS pairing $\hat\cH_P\otimes\hat\cH_{P'}\to\bC$,
but the induced intertwining operator
$\hat I_{P'P}\colon\hat\cH_P\to\hat\cH_{P'}$ is not necessarily unitary in
general.
Although we will not encounter this in our examples, we may need to augment
$\cH_P$ or $\hat\cH_P$ by a virtual sum of higher cohomology groups whose
elements are represented by differential forms valued in $\ell$ or
$\ell\otimes\sqrt{K_P}$.
In this way, when $M$ is compact, the dimensions of $\cH_P$ and $\hat\cH_P$ are
expected to stay constant as $P$ varies.

Suppose the phase space $(M,\om)$ and the prequantum line bundle $\ell$ remain
fixed but the polarisations are allowed to change in a parametrised family
$\cP$.
Except for symplectic vector spaces with linear polarisations and their
symplectic reductions or with a quite restrictive family of polarisations
(see \S\ref{lin-fix} and reference therein), there is no natural way to
identify, even projectively, the Hilbert spaces $\cH_P$ and $\cH_{P'}$ or
$\hat\cH_P$ and $\hat\cH_{P'}$ from two polarisations $P,P'\in\cP$ \cite{GM}.
The Hilbert spaces $\cH_P$ form a vector bundle $\cH$ over $\cP$.
When $\cP$ contains polarisations from almost complex structures only, $\cH$
is a sub-bundle of the product bundle whose fibre is the space of all
$L^2$-sections of $\ell$.
Therefore $\cH$ inherits a unitary connection $\nabla^\cH$.
With metaplectic correction, the bundle $\hat\cH\to\cP$ with fibres $\hat\cH_P$
also has a connection $\nabla^{\hat\cH}$.
The covariant derivative of a section $\varPsi$ of $\hat\cH$ at $P\in\cP$
along a tangent vector $S\in T_P\cP$ is defined by
\[ \bra(\nabla^{\hat\cH}_S\varPsi)(P),\psi'\ket_P
   =\mfrac\dr{\dr s}\Big|_{s=0}\bra\varPsi(P_s),\psi'\ket_{\mathrm{BKS}} \]
for any $\psi'\in\hat\cH_P$, where $P_s$ is a $1$-parameter family of
polarisations in $\cP$ with $P_0=P$ and $\frac{\dr P_s}{\dr s}\big|_{s=0}=S$.
The connection is unitary even though the operator $\hat I_{P'P}$ is not when
$P,P'\in\cP$ are of a finite distance apart.
When $P$ is joined to $P'$ by a path in $\cP$, we can identify $\cH_P$ with
$\cH_{P'}$ or $\hat\cH_P$ with $\hat\cH_{P'}$ by the parallel transport in
$\cH$ or $\hat\cH$ along the path.
But since the connections on $\cH$ and $\hat\cH$ are not projectively flat in
general, the identifications (up to a phase) of Hilbert spaces from different
polarisations are not canonical.

Given a classical observable $f\in C^\infty(M)$, the prequantum operator on
$\psi\in\Gam(\ell)$ is
\[ (-\ii\,\nabla_{X_f}+f)\,\psi=
   -\ii\mfrac\dr{\dr s}\Big|_{s=0}\tilde\phi_s^{-1}\circ\psi\circ\phi_s, \]
where $X_f$ is the Hamiltonian vector field of $f$, and the Hamiltonian flow
$\phi_s$ on $M$ lifts to $\tilde\phi_s$ on $\ell$ preserving $\nabla^\ell$.
For quantisation, we choose a polarisation $P$.
If the flow $\phi_s$ preserves $P$, then the above prequantum action preserves
the subspace $\Gam_{\bar P}(\ell)$ of sections parallel along $\bar P$ and
hence acts as an operator $O_P(f)$ on $\cH_P$;
otherwise, $\tilde\phi_s\circ\psi\circ\phi_s^{-1}$ is in
$\Gam_{\bar P_s}(\ell)$, where $P_s=(\phi_s)_*P$, and a projection onto
$\Gam_P(\ell)$ or $\cH_P$ should be inserted in the definition of $O_P(f)$.
With metaplectic correction, we use instead the lift $\hat\phi_s$ of $\phi_s$
on $\ell\otimes\sqrt{K_P}$.
The quantum operator of $f$ acting on $\psi\in\hat\cH_P$ is
\[ \hat O_P(f)\,\psi=-\ii\mfrac\dr{\dr s}\Big|_{s=0}\hat I_{PP_s}
                \big(\hat\phi_s^{-1}\circ\psi\circ\phi_s\big). \]
When polarisations vary in $\cP$, if the connection on the bundle of quantum
Hilbert space is flat or projectively flat, as in the case of symplectic vector
spaces and linear polarisations, then the bundle of operators on the Hilbert
spaces is flat, and we can expect that $O_P(f)$ or $\hat O_P(f)$ for various
$P\in\cP$ form a parallel section.
But little is known in general.
After fixing a suitable polarisation $P$, we denote the quantum operators
$O_P(f)$ or $\hat O_P(f)$ loosely by $\hat f$.

\subsection{Quantisation of a family of phase spaces}\label{gen-var}
Now suppose we have a family of symplectic phase spaces $(M_b,\om_b)$
parametrised by $b\in\cB$, where $\cB$ is a connected manifold containing the
parameters.
We assume that these $M_b$ form a space $M$ fibered over $\cB$ and that there
is a $2$-form $\om$ on $M$ whose restriction to each fibre $M_b$ over $b\in\cB$
is $\om_b$.
The total space $M$ has a vertical tangent bundle $\Tv M$: at each
$x\in M$ over $b\in\cB$, the fibre is $\Tv_xM=T_xM_b$.
The complement (with respect to $\om$) of $\Tv M$ in $TM$ is a
vector bundle $T^{\mathrm{h}}M$ over $M$ of rank $\dim\cB$ satisfying
$T^{\mathrm{h}}M\cap \Tv M=\{0\}$.
This follows from a simple argument in linear algebra: since
$\om|_{T_xM_b}=(\om_b)_x$ is non-degenerate on $T_xM_b$, the map
$T_xM\to(T_xM_b)^*$, $X\mapsto\iota_X\om_x|_{T_xM_b}$ is surjective and its
kernel $T^{\mathrm{h}}_xM$ is transverse to $T_xM_b$ and has dimension
$\dim T_xM-\dim(T_xM_b)^*=\dim\cB$.
The splitting $TM=\Tv M\oplus T^{\mathrm{h}}M$ defines a connection
of the fibration $M\to\cB$.
A vector $S\in T_b\cB$ on the parameter space $\cB$ can be lifted uniquely,
for any point $x\in M$ over $b$, to a horizontal vector $\tS\in T_xM$, and so
can a vector field on $\cB$.
Henceforth we suppose that $\om$ is closed.
Then as the parameter $b$ moves in $\cB$ along a prescribed curve $b=b(t)$,
the flow on $M$ along the horizontal lift $\tS_t$ of $S_t:=\dot b(t)$ generates
symplectomorphisms among the fibres $(M_{b(t)},\om_{b(t)})$.
Indeed, given two vertical vector fields $X$ and $Y$ on $M$, the bracket $[X,Y]$
remains vertical, and thus for any vector field $S$ on $\cB$ we have
\[ (\pounds_{\!\tS}\,\om)(X,Y)=(\dr\,\iota_{\!\tS}\,\om)(X,Y)
   =X(\om(\tS,Y))-Y(\om(\tS,X))-\om(\tS,[X,Y])=0.  \]

In the special case when $\om$ remains closed and has rank exactly equal to
$\dim M_b$ everywhere on $M$, horizontal vector fields $\tS$ on $M$ satisfy
$\iota_{\!\tS}\,\om=0$ and hence $\pounds_{\!\tS}\,\om=0$.
If so, for two vector fields $S$ and $T$ on $\cB$, the bracket $[\tS,\tilde T]$
of their horizontal lifts $\tS$ and $\tilde T$ remains horizontal because
$\iota_{[\tS,\tilde T]}\,\om=[\pounds_\tS,\iota_{\tilde T}]\,\om=0$.
Thus the connection on $M\to\cB$ is flat in this case.
However, we will not encounter examples of this special type below.
Instead, our total space will often be a product $M=M_0\times\cB$, although the
$2$-form $\om$ on $M$ will have various components $\om^{p,q}$, $p+q=2$,
according to the bi-grading of $2$-forms on $M_0\times\cB$, and $\om$ may have
a rank larger than the fibre dimension.
Then for a vector field $S$ on $\cB$, the horizontal lift to $M$ is
$\tS=(\tS-S)+S$, where the component $\tS-S$ along $M_0$ according to the
product structure is in general non-zero and is determined by the equation
$\iota_{\tS-S}\,\om^{2,0}=-\iota_S\,\om^{1,1}$.

Next, we consider prequantisation of the family given by a symplectic
fibration $M\to\cB$ as above.
A prequantum line bundle over the total space $M$ is a line bundle $\ell$ with
a unitary connection $\nabla^\ell$ whose curvature is $\om/\ii$.
To each fibre $M_b$, the restriction of $\ell$ is a prequantum line bundle
$\ell_b$ of the symplectic manifold $(M_b,\om_b)$.
As such its first Chern class $c_1(\ell_b)\in H^2(M_b,\bZ)$ modulo torsion is
represented in the de Rham cohomology by the closed $2$-form $\om_b/2\pi$.
Moreover, as $b$ varies continuously in $\cB$, the topology of $M_b$ and
$\ell_b$ remains unchanged.
That is, once a choice of $\ell_{b_0}$ is made for a particular $b_0\in\cB$,
then the Chern class $c_1(\ell_b)$, including its torsion part, for any other
$b\in\cB$ is also fixed.
There is still an ambiguity of flat line bundles that are topologically trivial
or of zero first Chern class.
But as in the case of a single phase space, they can be absorbed by adding to
the connection harmonic $1$-forms (modulo gauge transformations) that represent
the element in $H^1(M_b,\bZ)\otimes\U(1)$.
Finally, a prequantum line bundle over $M$ can be tensored by the pull-back of
a possibly topologically non-trivial flat line bundle over $\cB$.
In the following, we will assume that we have already made a choice of a
prequantum line bundle $\ell$ with connection $\nabla^\ell$ over $M$.

To quantise the family of phase spaces, we assume that polarisations $P_b$ on
$(M_b,\om_b)$, $b\in\cB$, can be chosen so that there is a complex distribution
$P$ on $M$ that restricts to the polarisation $P_b$ on each $(M_b,\om_b)$.
Then the quantum Hilbert spaces $\cH_b$ or $\hat\cH_b$, if their dimensions do
not jump as $b\in\cB$ varies, form a vector bundle $\cH$ or $\hat\cH$
(possibly of infinite rank) over the parameter space $\cB$.
Note that the bundles are now over the physical parameter space $\cB$ of the
systems rather than the space of polarisations.
In the best situations, as will happen in the examples considered below, $P$
is invariant under the flow of the horizontal lift $\tS$ on $M$ of any vector
field $S$ on $\cB$; this is the closest analogue of having a fixed
polarisation on a single phase space.
We argue that there are natural connections $\nabla^\cH$ and $\nabla^{\hat\cH}$
on the bundles $\cH$ and $\hat\cH$ of Hilbert spaces.
If a point $b_0\in\cB$ flows to another $b_1\in\cB$ along a vector field $S$
on $\cB$, we can parallel transport a section of $\ell$ or
$\ell\otimes\sqrt{K_P}$ on $M_{b_0}$ to one on $M_{b_1}$ along the horizontal
lift $\tS$ on $M$ using the connections on $\ell$ and on $\Tv M$.
In this procedure, a section parallel along $\bar P_{b_0}$ is transported to
a section parallel along $\bar P_{b_1}$.
Since the flow preserves $\om_b$ along the way, the connections defined on
$\cH$ and $\hat\cH$ are unitary.
If the flows along the horizontal vector fields $\tS$ do not preserve $P$,
then the transported section on $M_{b_1}$ is in the quantum Hilbert space in
a different polarisation and has to be converted to the Hilbert spaces in the
desired polarisations, after each infinitesimal step, using the projection
or the operator from the BKS pairing.
We then have a mixture of the effects from changing the parameter $b\in\cB$
of the classical systems and the polarisation in quantisation.

Suppose there is a family of Hamiltonian functions $H_b$ on $M_b$ that form a
smooth function $H$ on $M$.
The classical Hamiltonians $H_b$, upon quantisation, become operators
$\hat H_b$ acting on $\cH_b$ or $\hat\cH_b$, and they form a section of the
bundle of operators on $\cH$ or $\hat\cH$.
If $b$ moves in $\cB$ along the curve $b=b(t)$, then we have a quantum system
with a time-dependent Hamiltonian operator $\hat H_{b(t)}$.
Let $\nabla$ denote the connection $\nabla^\cH$ on $\cH$ or $\nabla^{\hat\cH}$
on $\hat\cH$ depending on whether metaplectic correction is used.
We postulate that the Schr\"odinger equation governing the quantum evolution is
\[ \ii\,\nabla_{\dot b(t)}\psi(t)=\hat H_{b(t)}\psi(t), \]
where $\psi(t)\in\cH_{b(t)}$ is the quantum state at time $t$.
The replacement of the ordinary time derivative of $\psi(t)$ by the covariant
derivative is a reflection of the fact that the symplectic structure and
prequantum line bundle change in the course of evolution.
If $\hat H_b=0$ for all $b$, then $\psi(t)$ evolves by a parallel transport in
$\cH$ or $\hat\cH$ along $b(t)$, and we have (non-Abelian) geometric phases when
$b$ undergoes cyclic changes in $\cB$.
However, if $P$ is not invariant under parallel transport of horizontal vector
fields on $M$, there is a mixture of the physical phase from the change of $b$
with the unphysical phase from the change of $P_b$.
A possible solution, which we will not pursue here, is to form a family of
Hilbert spaces over $\cB\times\cP$.
Then as $b=b(t)$ moves in $\cB$, we use the $1$-parameter family of
polarisations $P_t$ from the parallel transport in the vertical tangent
bundle $\Tv M$.

We assume that the energy spectrum of each $\hat H_b$ is discrete and bounded
from below, and is ordered such that $E_b^{(0)}<E_b^{(1)}<E_b^{(2)}<\cdots$.
We also assume that as $b\in\cB$ varies, the energy levels
$E_b^{(k)}<E_b^{(k+1)}$ do not cross.
Let $\cH_b^{(k)}$ be the eigenspace of $\hat H_b$ of energy $E_b^{(k)}$.
Then $\cH_b^{(k)}$ form a sub-bundle $\cH^{(k)}$ of $\cH$ and
$\cH=\bigoplus_{k\in\bN}\cH^{(k)}$.
There is a unitary connection $\nabla^{(k)}$ on $\cH^{(k)}$ as a sub-bundle of
$\cH$ with connection $\nabla^\cH$.
When the parameter $b\in\cB$ undergoes cyclic changes, the holonomies of
$\nabla^{(k)}$ are (non-Abelian) quantum adiabatic phases at the $k$th energy
level.
We further assume that the polarisations $P_b$ can be arranged so that the
sub-bundle $P\subset(\Tv M)^\bC$ is invariant under horizontal vector fields
lifted from $\cB$.
The holonomy along a loop $b(t)$ in $\cB$ has two contributions, one
from the non-triviality of the bundle $\cH$, counting the change of the phase
spaces $(M_{b(t)},\om_{b(t)})$ themselves and another from the sub-bundle
$\cH^{(k)}$ inside $\cH$, counting the change of the Hamiltonians
$\hat H_{b(t)}$.
If $M$ is a product $M_0\times\cB$ and if $\om_b$, $\ell_b$ and $P_b$ are
independent of $b\in\cB$, the bundle of Hilbert spaces $\cH$ is also a product
$\cH_0\times\cB$ with the trivial connection.
If the Hamiltonians $H_b$ do depend on $b\in\cB$, the $k$th eigenspaces
$\cH_b^{(k)}$ define a map from $\cB$ to the Grassmannian $\Gr(n_k,\cH_0)$
of $n_k$-planes in $\cH_0$,
where $n_k=\dim_\bC\cH_b^{(k)}$ is the degeneracy of $E_b^{(k)}$.
The connection $\nabla^{(k)}$ on $\cH^{(k)}$ is the pull-back of the universal
connection on $\Gr(n_k,\cH_0)$ \cite{Wu88}.
This gives the usual Berry's phases \cite{B84,Si,WZ} when the Hamiltonians
alone undergo cyclic changes.
The statements for the metaplectically corrected Hilbert space bundle $\hat\cH$
are identical.

\section{Quantisation of linear phase spaces}\label{lin}
\subsection{Phase spaces with fixed symplectic or Euclidean structures}
\label{lin-fix}
We first consider linear bosonic phase spaces.
Let $V$ be a finite dimensional vector space of dimension $2n$.
A linear symplectic structure on $V$ is a symplectic form that is translation
invariant and is thus an element $\om\in\medwedge^{\!2}\,V^*$.
Similarly, a linear complex structure on $V$ is a linear operator $J$ on $V$
such that $J^2=-\id_V$.
Let $\cJ(V,\om)$ be the space of compatible linear complex structures on
$(V,\om)$.
Each $J\in\cJ(V,\om)$ defines a positive complex polarisation on $(V,\om)$ and
determines a quantum Hilbert space $\cH_J$.
The collection $\{\cH_J\}$ forms a bundle $\cH$ of Hilbert spaces over
$\cJ(V,\om)$ which admits a projectively flat connection $\nabla^\cH$
\cite{ADPW}.
The base space $\cJ(V,\om)$ is a non-compact Hermitian symmetric space of
type $\Sp(2n,\bR)/\U(n)$, and the curvature of the projectively flat connection
$\nabla^\cH$ is $\upsig_\om\id_\cH/2\ii$, proportional to the standard K\"ahler
form $\upsig_\om$ on $\cJ(V,\om)$.
With metaplectic correction, the bundle of Hilbert spaces over $\cJ(V,\om)$ is
$\hat\cH=\cH\otimes\sqrt\cK$, where $\cK:=\medwedge^\mathrm{\!top\,}\cV^*$ and
$\cV$ is the tautological vector bundle whose fibre over $J\in\cJ(V,\om)$ is
$V^{1,0}_J$.
The bundle $\cV$ has a natural connection because $\cV\oplus\bar\cV$ is the
product bundle $V^\bC\times\cJ(V,\om)$ with the trivial connection.
The curvature of the natural connection on $\cK$ is $\ii\upsig_\om$, and thus
the connection $\nabla^{\hat\cH}$ on $\hat\cH$ is flat.
Real polarisations given by real Lagrangian subspaces of $(V,\om)$ lie on the
Shilov boundary of $\cJ(V,\om)$, and parallel transports in $\cH$ or $\hat\cH$
along geodesics in $\cJ(V,\om)$ from a real polarisation to a complex
polarisation or another real polarisation is the Segal-Bargmann or the Fourier
transform \cite{KW}.
(Strictly speaking, convergence of parallel transports in $\cH$ at the infinite
ends of geodesics requires a half-density correction which does change
projective flatness \cite{Wu11}.)

A similar analysis can be carried out for linear fermionic systems.
The phase space is an oriented Euclidean vector space $(V,g)$ of dimension $2n$,
and a polarisation is given by a linear complex structure $J$ on $V$ compatible
with the orientation on $V$ and the Euclidean metric $g$; the latter means that
$g(J\cdot,J\cdot)=g(\cdot,\cdot)$.
These complex structures form a space $\cJ(V,g)$ which can be identified with
a compact Hermitian symmetric space of type $\SO(2n)/\U(n)$ and has a standard
K\"ahler form $\upsig_g$.
For each $J\in\cJ(V,g)$, the quantum Hilbert space $\cH_J$ is, up to a crucial
fermionic exponential factor, the (unique) irreducible representation of the
Clifford algebra of $(V,g)$ constructed using $J$.
The collection $\{\cH_J\}$ forms a vector bundle $\cH$ over $\cJ(V,g)$ with a
projectively flat connection whose curvature is $\ii\upsig_g\id_\cH/2$ (see
\cite{Wu15}, but the definition of $\upsig_g$ has the opposite sign there).
Like in the bosonic case, there is a connection on the bundle $\cV$ over
$\cJ(V,g)$ whose fibre over $J$ is $V^{1,0}_J$.
The curvature of the induced connection on 
$\cK:=\medwedge^\mathrm{\!top\,}\cV^*$ is $\ii\upsig_g$.
Due to the opposite transformation rule of fermionic volume elements, fermionic
half-forms are in the opposite line bundle $\sqrt{\cK^{-1}}$.
The bundle of metaplectically corrected Hilbert spaces is therefore
$\hat\cH=\cH\otimes\sqrt{\cK^{-1}}$, and its connection $\nabla^{\hat\cH}$
is still flat \cite{Wu15}.
See \cite{Wu17} for a survey and for a discussion on odd dimensional fermionic
phase spaces.

We continue to consider a fixed linear bosonic phase space $(V,\om)$ from which
we obtain a quantum Hilbert space $\cH_0$ by using any fixed polarisation.
We now regard $J\in\cJ(V,\om)$ not as a polarisation but as a parameter in the
Hamiltonians $H_J:=\frac12\om(\cdot,J\cdot)$ of generalised harmonic
oscillators.
The decomposition according to the energy eigenspaces gives $\cH_0$ a Fock
space structure.
That is, $\cH_0=\bigoplus_{k\in\bN}\cH_J^{(k)}$, where for each
$k\in\bN=\{0,1,2,\dots\}$, $\cH_J^{(k)}=\mathrm{Sym}^k(V_J^{1,0})^*\,|0\ket_J$,
of dimension ${n+k-1\choose k}$, is the eigenspace of the Hamiltonian operator
$\hat H_J$ with energy $k+n/2$.
For each $k$, $\{\cH_J^{(k)}\}$ form a vector bundle $\cH^{(k)}$ over
$\cJ(V,\om)$ and there is a connection $\nabla^{(k)}$ on each $\cH^{(k)}$.
For $k=0$, the line bundle $\cH^{(0)}$ is spanned by the vacuum vectors
$|0\ket_J$ and has curvature $\ii\upsig_\om/2$.
For a general $k$, we have $\cH^{(k)}=\mathrm{Sym}^k\,\cV^*\otimes\cH^{(0)}$;
the connection $\nabla^{(k)}$ comes from the tensor product of the natural
connection on $\cV$ and that of $\cH^{(0)}$ and is not projectively flat if
$n>1$, $k\ge1$ \cite{Wu11}.
But if $n=1$, then for all $k\in\bN$, $\cH^{(k)}=(\cH^{(0)})^{\otimes(2k+1)}$
are line bundles and have curvatures $\ii(k+\frac12)\upsig_\om$; the factor
$k+\frac12$ matches \cite{B85}.
It is curious that for any $n\ge1$, the curvature of $\nabla^{(0)}$ is the
opposite of that of the projectively flat connection when polarisation
varies \cite{Wu11}.
This is because in the polarisation given by the same $J$, the vacuum wave
function $\exp(-H_J/2)$ of $\hat H_J$ is always real and therefore no phase
can arise as $J$ varies.
To avoid cancellation of the physical Berry's phase by the unphysical phase
from a change of polarisations, we should quantise with a fixed polarisation
or include metaplectic corrections (which works at least for linear phase
spaces).

A parallel pattern exists for linear fermionic systems \cite{Wu11}.
The phase space $(V,g)$ of dimension $2n$ is fixed and so is the quantum
Hilbert space $\cH_0$.
Instead, $J\in\cJ(V,g)$ parametrises the Hamiltonians
$H_J=\frac12g(J\cdot,\cdot)\in\medwedge^{\!2\,}V^*$ of fermionic harmonic
oscillators.
The Fock space decomposition is $\cH_0=\bigoplus_{0\le k\le n}\cH_J^{(k)}$,
where $\cH_J^{(k)}=\medwedge^k(V^{1,0}_J)^*\,|0\ket_J$, of dimension
$n\choose k$, is the eigenspace of $\hat H_J$ with energy $k-\frac n2$.
When $J\in\cJ(V,g)$ varies, we have vector bundles
$\cH^{(k)}=\medwedge^{\!k\,}\cV^*\otimes\cH^{(0)}$, $0\le k\le n$, with
connections $\nabla^{(k)}$, which yield (non-Abelian) Berry's phases if
$J$ undertakes cyclic changes.
The line bundle $\cH^{(0)}$ from the vacua $|0\ket_J$ is $\sqrt{\cK^{-1}}$,
and its curvature is again the opposite of that of the projectively flat
connection when polarisation varies.
Unless $k=0,n$, the connection $\nabla^{(k)}$ is not projectively flat.

\subsection{Phase spaces with varying symplectic or Euclidean structures}
\label{lin-var}
We now explore another family of generalised harmonic oscillators on a linear
bosonic phase space $V$ of dimension $2n$ but with varying symplectic
structures.
Suppose $g$ is a fixed Euclidean metric on $V$ and the Hamiltonian is
$H=\frac12g(\cdot,\cdot)$ is now fixed, but the symplectic form
$\om_J:=g(J\cdot,\cdot)$ is allowed to change with the parameter
$J\in\cJ(V,g)$.
Note that $\cJ(V,g)$ parametrises bosonic instead of fermionic systems here.
For each $J\in\cJ(V,g)$, we have $\om_J=\dr\,\tht_J$, where
$\tht_J=\frac12\iota_D\,\om_J$ and $D$ is the Euler vector field on $V$.
In linear coordinates $\{x^\mu\}$ on $V$, we have
$D=x^\mu\frac\pdr{\pdr x^\mu}$,
$\om_J=\frac12\om_{\mu\nu}(J)\dr x^\mu\wedge\dr x^\nu$ and
$\tht_J=\frac14\om_{\mu\nu}(J)(x^\mu\dr x^\nu-x^\nu\dr x^\mu)$, where
$\om_{\mu\nu}(J)=\om_J(\frac\pdr{\pdr x^\mu},\frac\pdr{\pdr x^\nu})$ are
constants on $V$ but depend on $J$.
Then $\om=\tdr\,\tht_J$, where $\tdr$ is the exterior differential on
$V\times\cJ(V,g)$, is a closed $2$-form whose restriction to each
$V\times\{J\}$ is $\om_J$.
We will discover that the ``best'' polarisation on each copy $V\times\{J\}$ is
given by $J$ itself.
Then there is a bundle $\cH$ (or $\hat\cH$ with half-form correction) of
Hilbert spaces over $\cJ(V,g)$ and each fibre $\cH_J$ (or $\hat\cH_J$) is the
quantisation of $(V,\om_J)$ in the polarisation $J\in\cJ(V,g)$.

Given $J\in\cJ(V,g)$, let $\{z^i\}_{1\le i\le n}$ be the linear complex
coordinates on $V$ corresponding to a linear basis $\{e_i\}_{1\le i\le n}$ of
$V_J^{1,0}$ and write $\bar z^{\bar i}=\overline{z^i}$.
The complex conjugates $\bar e_{\bar i}:=\overline{e_i}$, $1\le i\le n$, form
a complex linear basis of $V_J^{0,1}=\overline{V_J^{1,0}}$.
The metric is $g=g_{i\bar j}
(\dr z^i\otimes\dr\bar z^{\bar j}+\dr\bar z^{\bar j}\otimes\dr z^i)$ and the
symplectic form is $\om_J=\om_{i\bar j}\dr z^i\wedge\dr\bar z^{\bar j}$, where
$\om_{i\bar j}=\ii g_{i\bar j}$.
The projections onto $V_J^{1,0}$ and $V_J^{0,1}$ are $P=\frac12(\id_V-\ii J)$
and $\bar P=\frac12(\id_V+\ii J)$, respectively.
Now suppose there is an infinitesimal deformation $\del J$ of $J$ in $\cJ(V,g)$
or, more precisely, $\del J\in T_J\,\cJ(V,g)$.
Since $g(J\cdot,\del J\cdot)+g(\del J\cdot,J\cdot)=0$, $\del J$ is
antisymmetric, i.e.,
$(\del J)_{ij}:=(\del J)_i^{\;\bar k}g_{\bar kj}=-(\del J)_{ji}$ and
$(\del J)_{\bar i\bar j}:=(\del J)_{\bar i}^{\;k}g_{k\bar j}
=-(\del J)_{\bar j\bar i}$, and so are $\del P=\frac\ii2\del J$ and
$\del\bar P=-\frac\ii2\del J$.
To the first order, $V_{J+\del J}^{1,0}$ and $V_{J+\del J}^{0,1}$ have bases
$\{e_i+\del e_i\}_{1\le i\le n}$ and
$\{\bar e_{\bar i}+\del\bar e_{\bar i}\}_{1\le i\le n}$, respectively, where
$\del e_i=(\del P)_i^{\;\bar j}\bar e_{\bar j}
=-\frac\ii2(\del J)_i^{\;\bar j}\bar e_{\bar j}$ and 
$\del\bar e_{\bar i}=(\del\bar P)_{\bar i}^{\;j}e_j
=\frac\ii2(\del J)_{\bar i}^{\;j}e_j$.
In fact, the map $V_J^{1,0}\to V^{1,0}_{J+\del J}$, $e_i\mapsto e_i+\del e_i$
is the infinitesimal parallel transport along $\del J$ in the bundle
$\cV\to\cJ(V,g)$ under its natural connection \cite{Wu15}.
With the same variation $\del J$, the dual bases $\{e^i\}$ and
$\{\bar e^{\bar i}\}$ change by
$\del e^i=-\frac\ii2(\del J)_{\bar j}^{\ i}\,e^{\bar j}$ and
$\del e^{\bar i}=\frac\ii2(\del J)_{\!j}^{\;\bar i}\,e^j$.

The above infinitesimal parallel transport in $\cV$ along $\del J$ is the flow
on $V\times\cJ(V,g)$ generated by the vector
\[ \widetilde{\del J}:=\del J
   -\mfrac\ii2(\del J)_i^{\;\bar j}z^i\mfrac\pdr{\pdr\bar z^{\bar j}}
   +\mfrac\ii2(\del J)_{\bar i}^{\;j}z^{\bar i}\mfrac\pdr{\pdr z^j}.   \]
Clearly, it preserves the family of complex structures or polarisations
parametrised by $\cJ(V,g)$: as $J$ deforms to $J+\del J$ in $\cJ(V,g)$, the
flow on $V\times\cJ(V,g)$ along $\widetilde{\del J}$ changes the complex
structure $J$ on $V\times\{J\}$ to $J+\del J$ on $V\times\{J+\del J\}$.
Moreover, since $\iota_{\del J}\,\tht_J=0$ and
\[ \iota_{\del J}\,\om=\del(\tht_J)=\mfrac12(\del J)_{ij}\,z^i\dr z^j
   +\mfrac12(\del J)_{\bar i\bar j}\,z^{\bar i}\dr z^{\bar j}
   =-\iota_{\widetilde{\del J}-\del J}\,\om_J, \]
the vector $\widetilde{\del J}$ is indeed in the kernel of the paring $\om$
with the vertical vector fields.
We have thus verified the conditions on the family of symplectic structures
and polarisations on $V$ parametrised by $\cJ(V,g)$.
Consequently, the flow along $\widetilde{\del J}$ preserves the restriction
$\om_J$ of $\om$ to the vertical fibres $V\times\{J\}$.

We now proceed to find the connection and curvature of the bundle
of Hilbert spaces.
The prequantum line bundle $\ell$ over $V\times\cJ(V,g)$ is topologically
trivial.
Identifying its sections with complex valued functions, we find that the
connection on $\ell$ is $\nabla^\ell=\tdr+\tht_J/\ii$, whose curvature is
$\om/\ii$.
The quantum Hilbert space $\cH_J$ without metaplectic correction is the space
of $L^2$-functions $\psi$ on $V$ satisfying $\nabla^J_{\bar i}\,\psi=0$.
The collection $\{\cH_J\}$ forms a bundle $\cH$ of Hilbert spaces over
$\cJ(V,g)$.
Under an infinitesimal variation $\del J$ of $J\in\cJ(V,g)$, the parallel
transport in $\cH$ is $\psi\mapsto\psi+\del\psi$, where
\[ \del\psi=-\nabla^J_{\widetilde{\del J}-\del J}\psi
   =-\mfrac\ii2(\del J)_{\bar i}^{\;j}\bar z^{\bar i}\nabla^J_{\!j}\psi
   =-\mfrac\ii2(\del J)_{\bar i}^{\;j}\bar z^{\bar i}
   (\nabla^J_{\!j}+g_{j\bar k}\bar z^{\bar k})\psi. \]
Note that $[\nabla^J_{\bar i},\nabla^J_j]=-g_{\bar ij}$, or
$[\nabla^J_{\bar i},\nabla^J_{\!j}+g_{j\bar k}\bar z^{\bar k}]=0$.
Although $\nabla^J_{\bar i}(\del\psi) =-\frac\ii2(\del J)_{\bar i}^{\;j}
(\nabla^J_{\!j}+g_{j\bar k}\bar z^{\bar k})\psi$ is non-zero,
\[ \nabla^{J+\del J}_{\bar e_{\bar i}+\del\bar e_{\bar i}}(\psi+\del\psi)
   =\nabla^J_{\bar i}(\del\psi)+\mfrac\ii2(\del J)_{\bar i}^{\;j}\nabla^J_{\!j}
   \psi+\mfrac1{2\ii}(\del J)_{\bar j\bar i}\bar z^{\bar j}\psi=0 \]
holds to the first order, verifying the consistency of our definition.
The gauge potential of $\nabla^\cH$ on $\cJ(V,g)$ is the $1$-form $\del J
\mapsto-\del\psi=\frac\ii2(\del J)_{\bar i}^{\;j}\bar z^{\bar i}\nabla^J_{\!j}$
whose value at $J$ is an operator on $\cH_J$.

To identify the connection $\nabla^\cH$, we recall that the diagonalisation of
the quantised  Hamiltonian operator $\hat H$ yields the Fock space
decomposition $\cH_J=\bigoplus_{k\in\bN}\cH_J^{(k)}$, where
$\cH_J^{(k)}=\mathrm{Sym}^k(V_J^{1,0})^*\,|0\ket_J$ is the $k$th energy level.
Let $\cH^{(k)}$ be the bundle formed by $\cH_J^{(k)}$, $J\in\cJ(V,g)$.
The vacuum $|0\ket_J$ in the polarisation $J$ has a wave function $\exp(-H/2)$
which is independent of $J\in\cJ(V,g)$ and is a parallel section of $\cH^{(0)}$
or $\cH$.
The first excited level $\cH_J^{(1)}$ is spanned by wave functions of the form
$\psi^i(z)=z^i\exp(-H(z,\bar z)/2)$, $1\le i\le n$.
A simple calculation shows that $\psi^i$ changes by
$\del\psi^i=-\frac\ii2(\del J)_{\bar j}^{\ i}\,\overline{\psi^j}$
under an infinitesimal variation $\del J$, which agrees with how a dual base
vector $e^i$ changes.
Thus the connection $\nabla^\cH$ preserves the sub-bundle
$\cH^{(1)}\cong\cV^*$, and on $\cH^{(1)}$ it agrees with natural connection
on $\cV^*$.
Moreover, $\nabla^\cH$ preserves every sub-bundle $\cH^{(k)}$, which is
isomorphic to $\mathrm{Sym}^{k\,}\cV^*$, and $\nabla^\cH$ on $\cH^{(k)}$ agrees
with the naturally induced connection on $\mathrm{Sym}^{k\,}\cV^*$.
Note that projections onto $\cH^{(k)}$ are not needed in this case because the
connection $\nabla^\cH$ preserves the sub-bundles.

Since in this model the symplectic structure is not fixed and neither is the
polarisation, the notion of half-forms does depend on the parameters. 
Therefore metaplectic correction gives a different bundle of Hilbert spaces,
$\hat\cH=\cH\otimes\sqrt{\cK}$, over $\cJ(V,g)$, where
$\cK=\medwedge^\mathrm{\!top\,}\cV^*$.
Using the Fock space decomposition of $\hat\cH_J$, we obtain sub-bundles
$\hat\cH^{(k)}=\cH^{(k)}\otimes\sqrt\cK$ of $\hat\cH$ from the $k$th energy
level for $k\in\bN$.
The connection $\nabla^{\hat\cH}$ on $\hat\cH$ still preserves all sub-bundles
$\hat\cH^{(k)}$.
But the line bundle $\hat\cH^{(0)}=\sqrt\cK$ from the vacua has curvature
$\ii\upsig_g/2$ and is no longer flat.
Consequently, there are non-trivial geometric phases on the vacuum vector when
$J$ undergoes cyclic changes in $\cJ(V,g)$.
This is in contrast with the above result of trivial phases when metaplectic
correction was ignored.
The connections and (non-Abelian) geometric phases of the higher energy levels
$\hat\cH^{(k)}$, $k\ge1$, are also modified by $\sqrt\cK$.
Like in many other cases of geometric quantisation, we expect that metaplectic
corrections produce more realistic results. 

Finally, we consider fermionic harmonic oscillators with varying
phase space structures.
The phase space is a real vector space $V$ of even dimension $2n$.
The Hamiltonian $H=\frac12\om(\cdot,\cdot)$ is given by a linear symplectic
form $\om$ on $V$ and is fixed.
But the fermionic phase space structure given by the Euclidean metric
$g_J=\om(\cdot,J\cdot)$ is parametrised by complex structures $J\in\cJ(V,\om)$
compatible with $\om$; here positivity $\om(\cdot,J\cdot)>0$ is a requirement
of unitarity rather than the condition that energy is bounded from below
\cite{W91}.
For each $J\in\cJ(V,\om)$, we use $J$ itself as the polarisation.
This results in the quantum Hilbert spaces $\cH_J$ without metaplectic
correction and $\hat\cH_J$ with metaplectic correction, and they form bundles
$\cH$ and $\hat\cH\cong\cH\otimes\sqrt{\cK^{-1}}$, respectively, over
$\cJ(V,\om)$.
With the Hamiltonian operator $\hat H$, the Fock space structures yield bundle
decompositions $\cH=\bigoplus_{0\le k\le n}\cH^{(k)}$ and
$\hat\cH=\bigoplus_{0\le k\le n}\hat\cH^{(k)}$, where for each
$k=0,1,\dots,n$, $\cH_J^{(k)}\cong\medwedge^k\cV^*$ and
$\hat\cH_J^{(k)}\cong\medwedge^k\cV^*\otimes\sqrt{\cK^{-1}}$.
The connections $\nabla^\cH$ on $\cH$ and $\nabla^{\hat\cH}$ on $\hat\cH$
preserve the sub-bundles $\cH_J^{(k)}$ and $\hat\cH_J^{(k)}$, respectively,
and restrict to the natural connections induced from $\cV$.
In particular, the vacuum line bundle $\cH^{(0)}$ without metaplectic
correction is flat.
But with metaplectic correction, $\hat\cH^{(0)}\cong\sqrt{\cK^{-1}}$ is not
flat and has curvature $\upsig_\om/2\ii$, whereas the bundle with the top
occupation number, $\hat\cH^{(n)}\cong\sqrt\cK$, has the opposite curvature.

\subsection{Quaternionic and paraquaternionic phase spaces}\label{qua}
A quaternionic vector space $V$ is a module over the algebra of quaternions,
generated over $\bR$ by $i,j,k$ satisfying $i^2=j^2=k^2=-1$, $ij=-ji=k$, etc.
So $\dim_\bR V=2n$, where $n$ is even, and there are linear complex structures
$J_a$ ($a=1,2,3$) on $V$ satisfying
$J_aJ_b=-\del\ab\id_V+\,\eps\ab^{\;\;\;c}J_c$.
Here $\eps\ab^{\;\;\;c}=\eps_{abd}\del^{cd}$ and $\eps_{abc}$ is totally
anti-symmetric with $\eps_{123}=1$.
Moreover, for any point in the $2$-sphere
$S^2=\{\xi\in\bR^3:\del\ab\xi^a\xi^b=1\}$, we have a linear complex structure
$J_\xi:=\xi^aJ_a$ on $V$.
For future applications, we note the identity
\[ J_\xi\,J_\eta=-(\xi\cdot\eta)\id_V+\;(\xi\times\eta)^aJ_a \]
for all $\xi,\eta\in S^2$.
Let $g$ be a Euclidean metric on $V$ that is invariant under all $J_a$
($a=1,2,3$) and hence under $J_\xi$ ($\xi\in S^2$).
Then we have symplectic forms $\om_a:=g(J_a\cdot,\cdot)$ and, more generally,
$\om_\xi:=g(J_\xi\cdot,\cdot)$ for all $\xi\in S^2$.
We consider the quantisation of the family of bosonic phase spaces
$(V,\om_\xi)$ parametrised by $\xi\in S^2$.

The solution to this problem comes readily when we apply the results in
\S\ref{lin-var}.
Indeed, since $g$ is fixed, the map $\xi\mapsto J_\xi$ embeds $S^2$ in
$\cJ(V,g)$.
(When $n=2$, the map is one-to-one.)
For each $\xi\in S^2$, we use the polarisation given by $J_\xi$ itself.
Then the bundle $\hat\cH$ of quantum Hilbert spaces over $\cJ(V,g)$ restricts
to $S^2$, and so does the Fock space decomposition
$\hat\cH=\bigoplus_{k\in\bN}\hat\cH^{(k)}$ that is preserved by the connection
$\nabla^{\hat\cH}$.
Explicitly, the K\"ahler form $\upsig_g$ on $\cJ(V,g)$ is
\[ \upsig_g=-\mfrac\ii4\tr_{V_J^{1,0}}(\dr J\wedge\dr J).  \]
Restricting to $S^2$, we have $\dr J_\xi=J_a\dr\xi^a$.
Using $\tr_V(J_\xi)=0$ and the formula for the product $J_\xi J_\eta$, we get
\[ \upsig_g
   =-\mfrac\ii4\tr_V\Big(\mfrac{1-\ii J_\xi}2J_aJ_b\Big)\dr\xi^a\wedge\dr\xi^b
   =\mfrac n2\,\upsig_{S^2}, \]
where $\upsig_{S^2}:=\frac12\,\eps_{abc}\,\xi^a\dr\xi^b\wedge\dr\xi^c$ is the
standard area form on $S^2$ such that $\int_{S^2}\upsig_{s^2}=4\pi$.
Consider the vacuum line bundle $\hat\cH^{(0)}\cong\sqrt\cK$ whose curvature
is $\frac\ii2\upsig_g=\frac{\ii\,n}4\upsig_{S^2}$.
When $\xi\in S^2$ undergoes a cyclic change, the geometric phase acquired by
the vacuum state is $\exp\big(\!-\!\frac\ii4n\Om\,\big)$, where $\Om$ is the
solid angle spanned by the loop in $S^2$.
Note that $\Om$ is defined modulo $4\pi$ and so the phase is well defined
because $n$ is even.

For the fermionic counterpart, we consider a paraquaternionic vector space $V$
which is a module over the algebra of split quaternions generated over $\bR$ by
$i,j,k$ with $-i^2=j^2=k^2=1$, $ij=-ji=-k$, $ik=-ki=j$, $jk=-kj=i$ \cite{L}.
So there are linear operators $J_a$ ($a=0,1,2$) on $V$ satisfying
$J_aJ_b=-\gam\ab\id_V+\,\eps\ab^{\;\;\;c}J_c$.
Here $(\gam\ab)=\mathrm{diag}(1,-1,-1)$ is the Minkowski metric on $\bR^{1+2}$,
$\eps\ab^{\;\;\;c}=\eps_{abd}\gam^{cd}$ and $\eps_{abc}$ is totally
anti-symmetric with $\eps_{012}=1$.
Among $J_a$ ($a=0,1,2$), $J_0$ is a linear complex structure, and so are
$J_\xi=\xi^aJ_a$ for all $\xi\in\bR^{1+2}$ on the $2$-sheeted hyperboloid
$\gam\ab\xi^a\xi^b=1$; the two sheets give opposite orientations on $V$.
We let $H^2:=\{\xi\in\bR^{1+2}:\xi^0>0,\gam\ab\xi^a\xi^b=1\}$.
The Minkowski metric restricts to a space-like negatively curved metric on
$H^2$ with the area form
$\upsig_{H^2}=\frac12\,\eps_{abc}\,\xi^a\dr\xi^b\wedge\dr\xi^c$.
So $H^2$ is the standard hyperbolic plane.
The real dimension of $V$ with a paraquaternionic structure is always even,
say $2n$, but is not necessarily a multiple of $4$.
The simplest realisation is $J_0=-\ii\sig_2$, $J_1=\sig_1$, $J_2=\sig_3$ on
$\bR^2$, where $\sig_1,\sig_2,\sig_3$ are the standard Pauli matrices.

We fix a symplectic form $\om$ on the paraquaternionic vector space $V$ that
is invariant under all $J_a$ and such that $\om(\cdot,J_0\cdot)$ is positive
definite.
Then for all $\xi\in H^2$, $\om$ is also invariant under the complex structure
$J_\xi$ and $g_\xi:=\om(\cdot,J_\xi\cdot)$ is also positive definite.
That is, $J_\xi\in\cJ(V,\om)$ and we have a family of fermionic phase spaces
$(V,g_\xi)$ parametrised by $\xi\in H^2$.
We use the embedding of $H^2$ in $\cJ(V,\om)$ given by $\xi\mapsto J_\xi$.
(The map is one-to-one if $n=1$.)
By a similar calculation, the restriction of the K\"ahler form $\upsig_\om$
on $\cJ(V,\om)$ to $H^2$ is
\[ \upsig_\om=-\mfrac\ii4\tr_{V_J^{1,0}}(\dr J\wedge\dr J)
             =\mfrac n2\,\upsig_{H^2}.  \]
As in the bosonic case, the Fock space decomposition
$\hat\cH=\bigoplus_{0\le k\le n}\hat\cH^{(k)}$ restricts to $H^2$.
In particular, the curvature of the vacuum line bundle
$\hat\cH^{(0)}\cong\sqrt{\cK^{-1}}$ is $-\frac\ii4n\upsig_{H^2}$.
The geometric phase acquired by the vacuum state when $\xi$ undergoes a cyclic
change is $\exp\big(\frac\ii4n\Om\,\big)$, where $\Om$ is the area bounded by
the loop in $H^2$.
The geometric phase on the state with the full occupation number is opposite.

\section{Conclusions}
After presenting the general theory of geometric quantisation of a family of
symplectic phase spaces, we specialise to the case of vector spaces.
The advantages are at least three-fold.
First, symplectic vector spaces are among few known examples of phase spaces
whose quantisation does not depend on the choice of polarisations.
In this way, it is possible to extract the physical geometric phases due to
the changes of Hamiltonians and symplectic structures without the mixture of
unphysical phases due to changes of polarisations.
Second, symplectic vector spaces have fermionic counterparts which are
oriented Euclidean vector spaces.
Other than sporadic examples such as reductions from linear fermionic phase
spaces \cite{Wu15}, there seems to be no systematic discussion on curved
fermionic spaces.
Third, linear phase spaces, both bosonic and fermionic, are cases in which
all structures can be worked out explicitly.
They provide crucial test grounds of the general theory and serve as important
examples for further developments.

Quaternionic vector spaces with invariant metrics generalise obviously to
hyper-K\"ahler manifolds.
On a hyper-K\"ahler manifold $M_0$, there is also an $S^2$-family of
symplectic forms.
Quantisation of the family requires using the total space $M=M_0\times S^2$,
which plays a key role in twistor theory (see \cite{DS} for a survey).
Equally interesting is the fermionic analogue: paraquaternionic vector spaces
with invariant symplectic forms.
The curved generalisation is known as a paraquaternionic K\"ahler manifold
(also denoted by $M_0$).
The total space $M=M_0\times H^2$ of a family appears in the hyperbolic
version of twistors (see the survey article \cite{IMZ}).
Many intriguing relations remain to be revealed.

\section*{Acknowledgments}
The author is supported in part by grant No.\;108-2115-M-007-004-MY2 from MoST,
Taiwan.

\bigskip
\makeatletter
\renewcommand\@biblabel[1]{#1.}
\makeatother

\end{document}